\documentclass[5p]{elsarticle}

\usepackage{hyperref}

\usepackage{lineno}
\modulolinenumbers[5]

\usepackage{amsmath}
\usepackage{amssymb}

\usepackage{graphicx}

\journal{Physics Letters A}

\begin{document}

\begin{frontmatter}

\title{Probing the gravitational redshift with an Earth-orbiting satellite}



\author[sai,asc,bmstu]{D.~A.~Litvinov\corref{mycorrespondingauthor}}
\cortext[mycorrespondingauthor]{Corresponding author at: Sternberg Astronomical Institute, Lomonosov Moscow State University,
Universitetsky pr.~13, 119991 Moscow, Russia.}
\ead{litvirq@yandex.ru}

\author[sai]{V.~N.~Rudenko\corref{mycorrespondingauthor}}
\ead{rvn@sai.msu.ru}

\author[asc]{A.~V.~Alakoz}

\author[mpifr]{U.~Bach}

\author[yorku]{N.~Bartel}

\author[sai]{A.~V.~Belonenko}

\author[asc]{K.~G.~Belousov}

\author[yorku,hart]{M.~Bietenholz}

\author[asc]{A.~V.~Biriukov}

\author[yarr]{R.~Carman}

\author[jive,astron]{G.~Cim\'o}

\author[azur]{C.~Courde}

\author[tudelft]{D.~Dirkx}

\author[caltech]{D.~A.~Duev}

\author[sai]{A.~I.~Filetkin}

\author[tudelft]{G.~Granato}

\author[jive,tudelft]{L.~I.~Gurvits}

\author[sai]{A.~V.~Gusev}

\author[chalmers]{R.~Haas}

\author[wettzell]{G.~Herold}

\author[utas]{A.~Kahlon}

\author[asc]{B.~Z.~Kanevsky}

\author[asc,bmstu]{V.~L.~Kauts}

\author[asc]{G.~D.~Kopelyansky}

\author[prao]{A.~V.~Kovalenko}

\author[bkg]{G.~Kronschnabl}

\author[sai]{V.~V.~Kulagin}

\author[asc]{A.~M.~Kutkin}

\author[chalmers]{M.~Lindqvist}

\author[utas]{J.~E.~J.~Lovell}

\author[azur]{H.~Mariey}

\author[utas]{J.~McCallum}

\author[jive,fgri]{G.~Molera~Calv\'es}

\author[mstromlo]{C.~Moore}

\author[yorku]{K.~Moore}

\author[tum]{A.~Neidhardt}

\author[bkg]{C.~Pl\"otz}

\author[jive]{S.~V.~Pogrebenko}

\author[mstromlo]{A.~Pollard}

\author[sai,mpifr]{N.~K.~Porayko}

\author[hart]{J.~Quick}

\author[asc]{A.~I.~Smirnov}

\author[sai,asc,athens]{K.~V.~Sokolovsky}

\author[kiam]{V.~A.~Stepanyants}

\author[azur]{J.-M.~Torre}

\author[yebes]{P.~de~Vicente}

\author[chalmers]{J.~Yang}

\author[kiam]{M.~V.~Zakhvatkin}

\address[sai]{Sternberg Astronomical Institute, Lomonosov Moscow State University,
Universitetsky pr.~13, 119991 Moscow, Russia}

\address[asc]{Astro Space Center, Lebedev Physical Institute, Profsoyuznaya 84/32, 117997 Moscow, Russia}

\address[bmstu]{Bauman Moscow State Technical University, 2-ya Baumanskaya 5, 105005 Moscow, Russia}

\address[mpifr]{Max-Planck-Institut f\"ur Radioastronomie, Auf dem H\"ugel 69, 53121 Bonn, Germany}

\address[yorku]{York University, Toronto, Ontario M3J 1P3, Canada}

\address[hart]{Hartebeesthoek Radio Astronomy Observatory, P.O. Box 443, Krugersdorp 1740, South Africa}

\address[yarr]{Yarragadee Geodetic Observatory, Geoscience Australia, PO Box 137 Dongara 6525, Western Australia}

\address[jive]{Joint Institute for VLBI ERIC, PO Box 2, 7990 AA Dwingeloo, The Netherlands}

\address[astron]{ASTRON, the Netherlands Institute for Radio Astronomy, PO 2, 7990 AA Dwingeloo, The~Netherlands}

\address[azur]{Universit\'e C\^ote d'Azur, CNRS, Observatoire de la C\^ote d'Azur, IRD, G\'eoazur, 2130 route de l'Observatoire, 06460 Caussols, France}

\address[tudelft]{Department of Astrodynamics and Space Missions, Delft University of Technology, 2629~HS~Delft, The Netherlands}

\address[caltech]{California Institute of Technology, Pasadena, CA 91125, USA}

\address[chalmers]{Department of Space, Earth and Environment, Chalmers University of Technology, Onsala Space Observatory, 439 92 Onsala, Sweden}

\address[wettzell]{Geodetic Observatory Wetzell (GOW), Germany}

\address[utas]{School of Physical Sciences, University of Tasmania, Private Bag 37, 7001 Hobart, Australia}

\address[prao]{Pushchino Radio Astronomy Observatory, Astro Space Center, Lebedev Physical Institute, 142290 Pushchino, Russia}

\address[bkg]{Federal Agency for Cartography and Geodesy, Sackenrieder Str.~25, D-93444 Bad K\"otzting, Germany}

\address[fgri]{Finnish Geospatial Research Institute, Geodeetinrinne 2, 02430 Masala, Finland}

\address[mstromlo]{EOS Space Systems Pty Limited, Mt Stromlo Observatory, Cotter Road, Weston Creek, ACT 2611, Australia}

\address[tum]{Technical University of Munich, Geodetic Observatory Wettzell, Sackenrieder Str.~25, D-93444 Bad K\"otzting, Germany}

\address[athens]{IAASARS, National Observatory of Athens, Vas.~Pavlou \& I.~Metaxa, 15236~Penteli, Greece}

\address[kiam]{Keldysh Institute for Applied Mathematics, Russian Academy of Sciences, Miusskaya sq. 4, 125047 Moscow, Russia}

\address[yebes]{Observatorio de Yebes (IGN), Apartado 148, 19180 Yebes, Spain}

\begin{abstract}
We present an approach to testing the gravitational redshift effect using the RadioAstron satellite. The experiment is based on a modification of the Gravity Probe A scheme of nonrelativistic Doppler compensation and benefits from the highly eccentric orbit and ultra-stable atomic hydrogen maser frequency standard of the RadioAstron satellite. Using the presented techniques we expect to reach an accuracy of the gravitational redshift test of order $10^{-5}$, a magnitude better than that of Gravity Probe A. Data processing is ongoing, our preliminary results agree with the validity of the Einstein Equivalence Principle.
\end{abstract}


\end{frontmatter}


\section{Introduction}

Quantum theory and general relativity are the two pillars of modern physics. However, they are incompatible. Attempts to quantize gravitation in the frameworks
of string theory or loop quantum gravity inevitably lead to a violation of
the Einstein Equivalence Principle (EEP) and thus to a breakdown of the metric nature of gravitation \cite{will-2014-lrr}.
Although there exist attempts to preserve the unquantized status of gravitation, they have not created a compelling case so far \cite{carlip-2008-cqg}. 
Tests of the EEP are therefore
of primary interest to characterize any unified theory of interactions.

Much progress in the field of EEP tests was made with direct involvement and under the leadership of V.~B.~Braginsky, the founder of a gravitational physics school (see, e.g. \cite{braginsky-panov-1972-jetp,braginsky-rudenko-1978-phys-rep,braginsky-rudenko-1970-spu}). With many coauthors of this paper proudly regarding Vladimir Borisovich as a teacher, we would like to dedicate this work to his memory.

In this paper we present an approach
to probing the gravitational redshift effect, which constitutes a test of the Local Position Invariance aspect of the EEP, using the RadioAstron satellite
\cite{kardashev-2013-ar}.
A highly eccentric orbit  and an ultra-stable on-board atomic hydrogen maser
 clock   make RadioAstron a unique laboratory for such test.
The idea of experiments of this kind is to compare the rate of time flow
at different space-time points against
the gravitational potential difference between them. In the
simplest case when time is measured by identical clocks and the gravitational field is weak, the basic
equation reads \cite{will-2014-lrr}:
\begin{equation}
   {\Delta T\over T} = (1+\varepsilon) { \Delta U \over c^{2} },
\label{eq:main-time}  
\end{equation}
where $\Delta T / T$ is the fractional difference of time intervals
measured by the clocks, $\Delta U$ is the gravitational potential difference
between them, $c$ is the speed of light, and $\varepsilon$ is the violation
parameter to be determined. In unified theories $\varepsilon$ is usually non-zero and depends on the clock type and element
composition of the gravitational field source, while in general relativity and any other metric theory
of gravitation $\varepsilon = 0$.

The concept of a satellite-based gravitational redshift experiment was developed and realized by Vessot et al. in the suborbital Gravity Probe A (GP-A) mission \cite{vessot-levine-1980-prl}, which yielded the best such test to date: it found $\varepsilon = (0.05 \pm 1.4)\times 10^{-4}$ (for hydrogen maser clocks), with $\delta\varepsilon = 1.4\times 10^{-4}$ constituting the accuracy of the test (1$\sigma$). A modified
approach we have developed for RadioAstron allows us, in principle, to reach an accuracy of
$\delta\varepsilon \sim 10^{-6}$, benefitting from a better performing hydrogen maser (H-maser) and prolonged data accumulation \cite{biriukov-2014-ar}. However, technical and operational constraints discussed below
limit the achievable accuracy to $\delta\varepsilon \sim 10^{-5}$. Several competing experiments are currently at various stages of preparation or realization, with accuracy goals ranging from $4\times10^{-5}$ to $2\times10^{-6}$ \cite{delva-2015-cqg,aces-2011-acau,jetzer-2017-ijmpd}.

The RadioAstron project is an international collaborative mission centered around the 10-m space radio telescope, with the primary goal of performing Space VLBI (Very-Long-Baseline Interferometry) observations of celestial radio sources of different nature with an extraordinary high angular resolution \cite{kardashev-2013-ar}. 
The RadioAstron spacecraft is on a highly eccentric orbit around
the Earth, evolving due to the gravitational influence of the Moon, as well as other factors, within a broad range of the orbital parameter space (perigee altitude 1,000 -- 80,000~km, apogee altitude 270,000 -- 370,000~km). The gravitational redshift experiment is a part of the RadioAstron mission's Key Science Programme. The essential characteristic of the mission, making it suitable for the experiment, is the presence of the space-qualified H-maser VCH-1010 aboard the spacecraft \cite{vremya-ch-vch-1010}.

The outline of the paper is as follows. In Section 2 we present our approach
to testing the gravitational redshift effect with RadioAstron, emphasizing
similarities and differences between our Doppler compensation scheme and that of Gravity Probe A. In Section 3 we briefly discuss our data processing algorithms and describe how we treat small effects that are not cancelled by the Doppler
compensation
scheme. In Section 4 we give details of the measurements performed so far. We conclude with Section
5 by discussing the preliminary results and prospects for future research.

\section{Outline of the RadioAstron gravitational redshift experiment}

\begin{figure}[t]
\includegraphics[width=\linewidth]{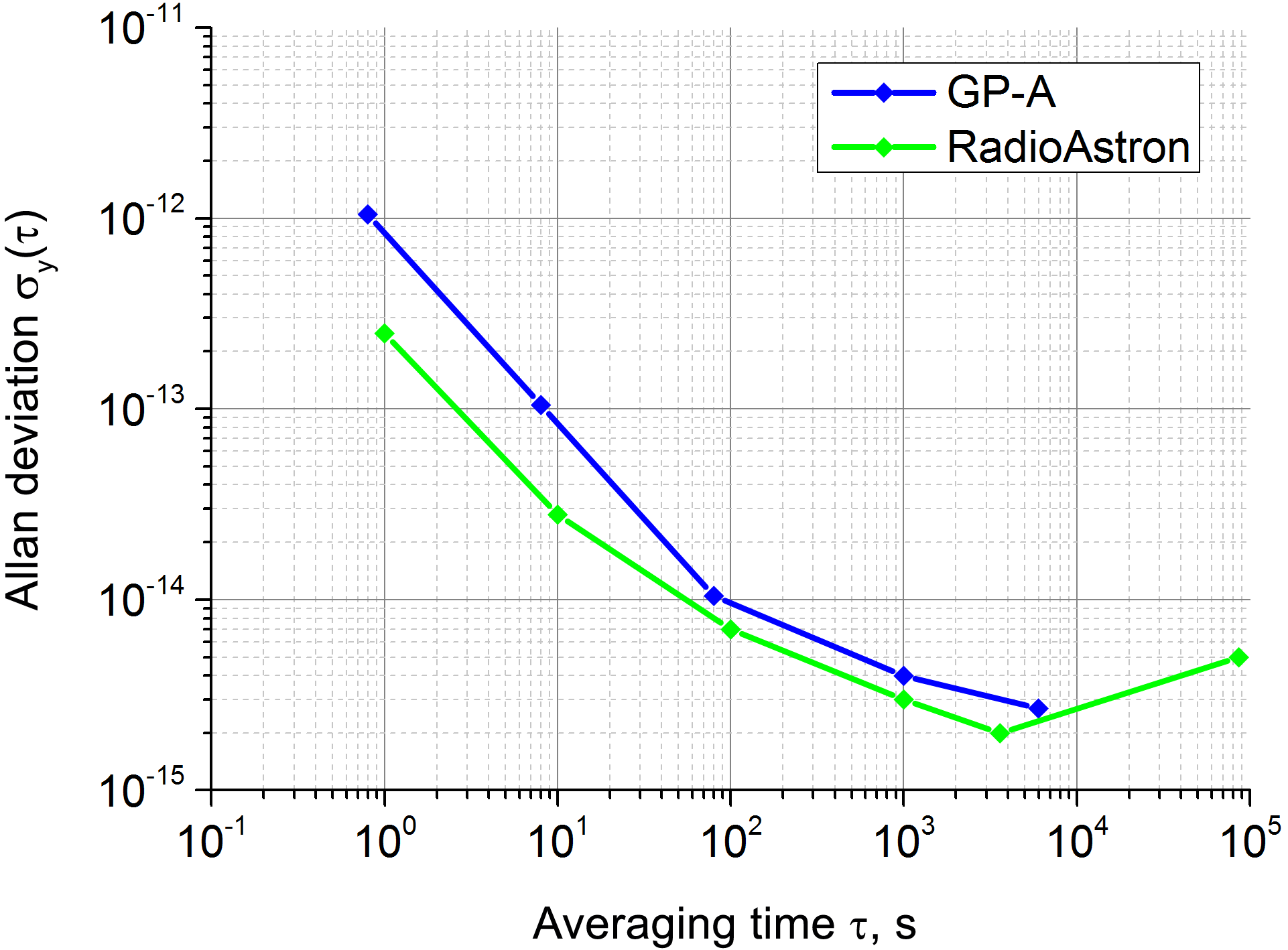}
\caption{Comparison of the frequency stability of the RadioAstron VCH-1010 \cite{vremya-ch-vch-1010} and GP-A VLG-10 \cite{smarr-vessot-1983-grg} H-masers in terms of the Allan deviation.}
\label{fig:adev}
\end{figure}

\begin{figure}[t]
\includegraphics[width=\linewidth]{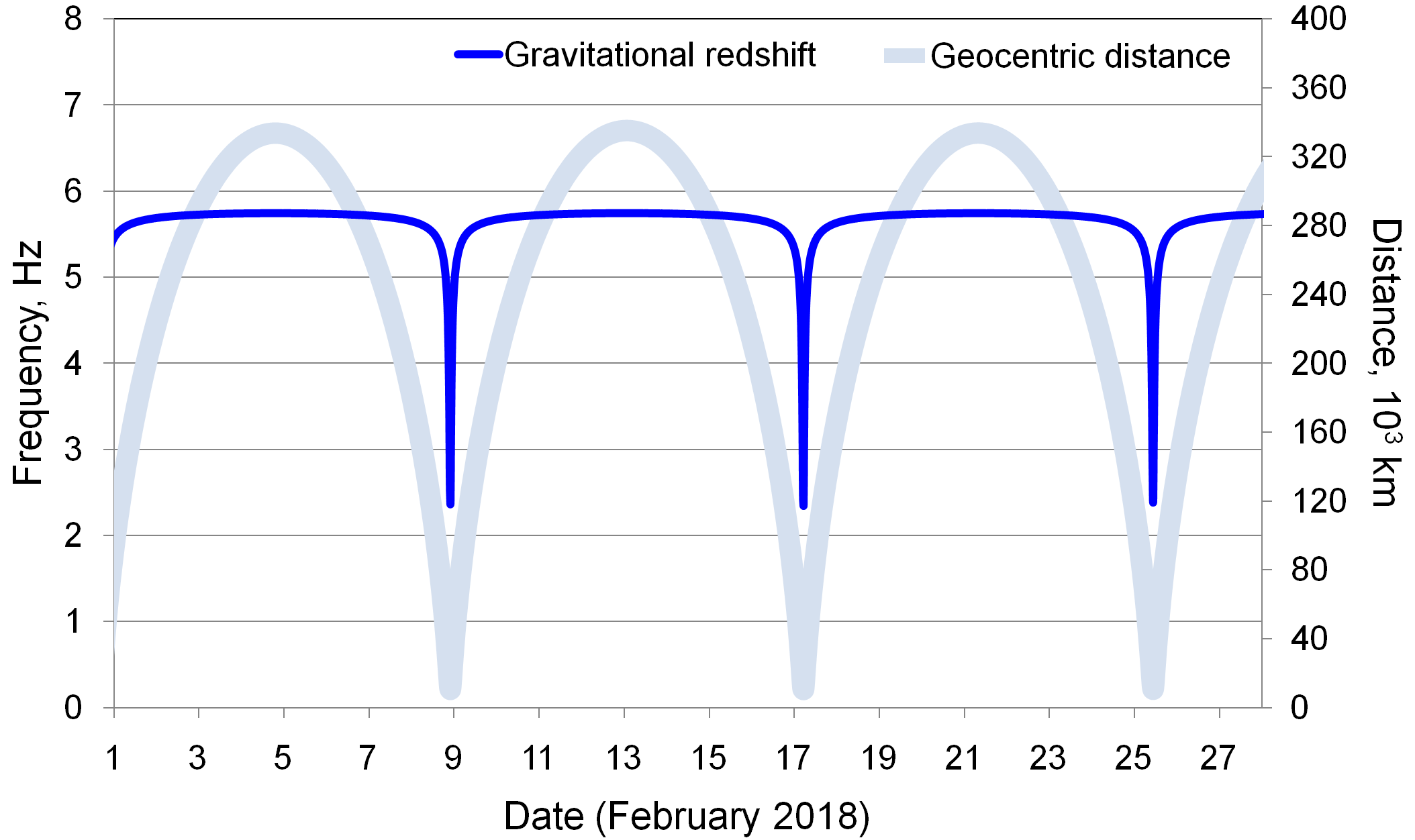}
\caption{Variation of the gravitational frequency shift of the 8.4~GHz downlink signal along the orbit during a low perigee epoch.}
\label{fig:redshift-modulation}
\end{figure}

There exist two approaches to testing the gravitational redshift effect in the field of the Earth. The first one is based on measuring the total value of the gravitational redshift between a ground station and a satellite. This approach, to be pursued by the ACES mission \cite{aces-2011-acau} and often called the absolute gravitational redshift measurement, is feasible only with  accurate clocks. The second approach, pioneered by Gravity Probe A, requires a stable clock, such as an H-maser, and is based on measuring the modulation of the redshift effect caused by the spacecraft's motion along an eccentric orbit around the Earth. We follow the second approach, benefiting from the high stability of RadioAstron's H-maser (Fig.~\ref{fig:adev}) and the deep modulation
of the redshift effect due to the high eccentricity of the orbit (Fig.~\ref{fig:redshift-modulation}). The modulation approach has an important advantage over the absolute measurement---it eliminates most systematic errors and provides for statistical averaging of the results. 
The ultimate goal of the experiment, in either case, is to determine the EEP violation parameter $\varepsilon$ by comparing the experimentally measured redshift, $\Delta T_\mathrm{grav}$ or $\Delta f_\mathrm{grav}$, against the computed gravitational potential difference, $\Delta U$, between the ground and space-borne clocks using Eq.~\eqref{eq:main-time} or \eqref{eq:main-freq} (below).

In the gravitational redshift experiment with RadioAstron we detect the frequency change of RadioAstron's on-board H-maser due to gravitation by comparing it, with the help of radio links,  with an H-maser at a
ground station. The fractional frequency shift due to gravitation, ${\Delta f_\mathrm{grav} / f}$, of a signal at frequency $f$ sent from the spacecraft to a ground station, is:
\begin{equation}
   {\Delta f_\mathrm{grav} \over f} = (1+\varepsilon) { \Delta U \over c^{2} },
\label{eq:main-freq}  
\end{equation}
which reflects the same physics as Eq.~\eqref{eq:main-time}.
Either one of RadioAstron mission's dedicated tracking stations (TS),
Pushchino (Moscow region, Russia) or Green Bank (West Virginia, USA), or a regular ground radio telescope (GRT) equipped with a
8.4 or 15 GHz receiver
 may be used to receive the spacecraft signal. The small gravitational frequency shift, with a maximum value of $\Delta f_\mathrm{grav}/f \sim 7\times10^{-10}$ at the apogee, needs to be extracted from a number of other effects influencing the signal sent from the spacecraft to the ground station \cite{vessot-levine-1979-grg}:

\begin{multline}
{\Delta f_{1\mathrm{w}}} =
\\
=
f \left(
 - {\dot D \over c} 
- {v_\mathrm{s}^2 - v_\mathrm{e}^2 \over 2 c^2}
+ { (\mathbf v_\mathrm{s} \cdot \mathbf n)^2
- (\mathbf v_\mathrm{e} \cdot \mathbf n) \cdot (\mathbf v_\mathrm{s} \cdot \mathbf n)
  \over c^2 }
\right)
\\
+ {\Delta f_\mathrm{grav} }
+ {\Delta f_\mathrm{ion} }
+ {\Delta f_\mathrm{trop} }
+ {\Delta f_\mathrm{fine} }
+ {\Delta f_0 }
+ O\left({v\over c}\right)^3,
\label{eq:one-way}
\end{multline}
where $\mathbf v_\mathrm{s}$ and $\mathbf v_\mathrm{e}$ are the velocities of the spacecraft and
the ground station (in a geocentric inertial reference
frame), $\dot D$ is the radial velocity of the
spacecraft relative to the ground station, $\mathbf n$ is a unit vector in the direction opposite to that of signal propagation, $\Delta f_\mathrm{ion}$  and $\Delta f_\mathrm{trop}$ are the ionospheric and tropospheric shifts, $\Delta f_\mathrm{fine}$ denotes
various fine effects (phase center motion, instrumental, etc.), $\Delta f_0$ is the frequency bias
between the ground and space H-masers, and ``$\mathrm{1w}$'' stands for ``1-way'' (space
to ground).

\begin{figure}[t]
\includegraphics[width=\linewidth]{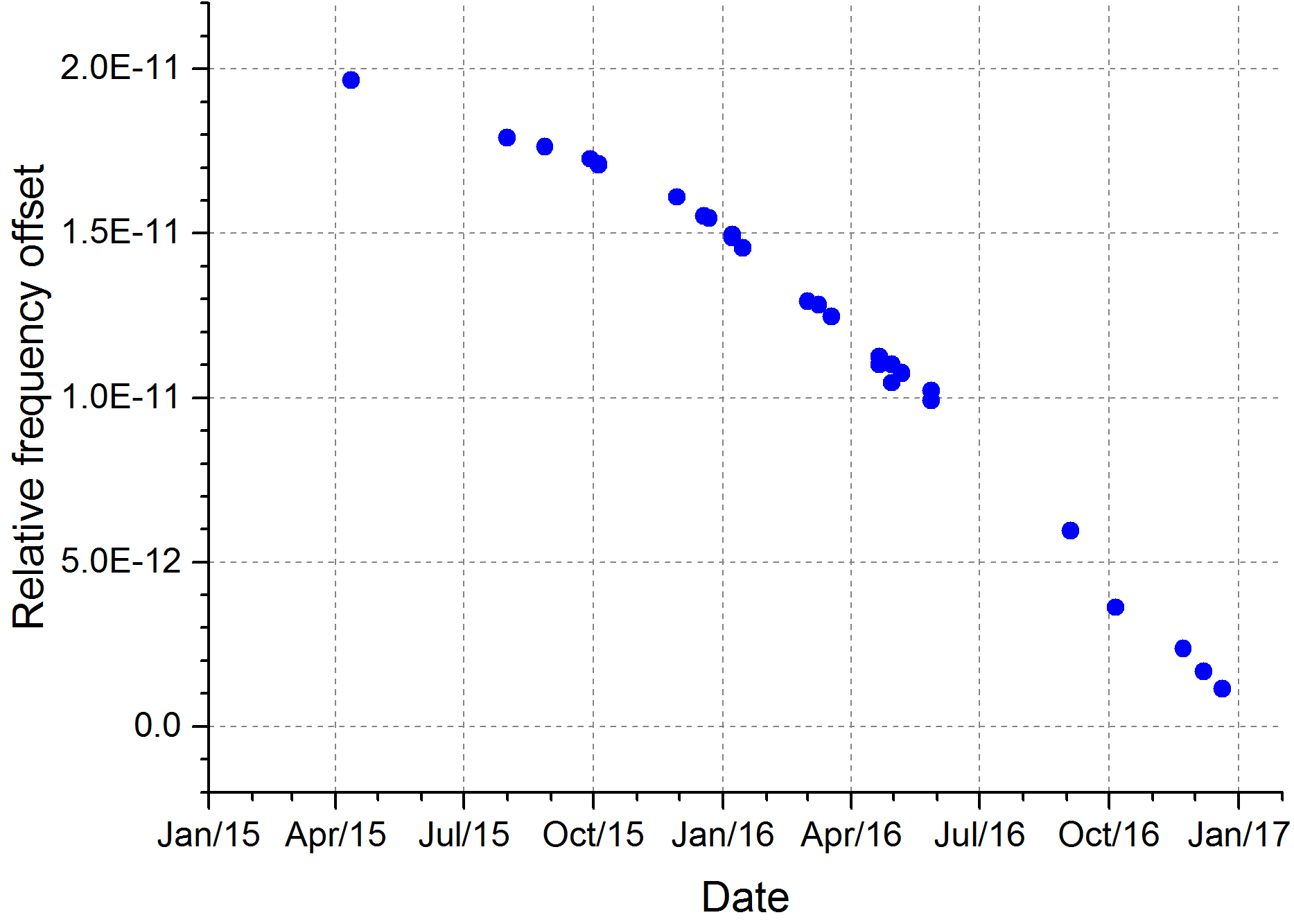}
\caption{Frequency drift of the RadioAstron on-board \mbox{H-maser} relative to the H-maser at the Green Bank TS.}
\label{fig:drift}
\end{figure}

There are two major problems in using Eq.~\eqref{eq:one-way} to determine $\Delta f_\mathrm{grav}$ directly, at
least for RadioAstron. First, the frequency bias, $\Delta f_0$, cannot be determined after launch without making
use of Eq.~\eqref{eq:main-freq}, which requires the knowledge of $\varepsilon$. We do not expect $\Delta f_0$ to be negligible for H-maser standards and, moreover, the
well-known cavity ``pulling'' effect may cause it to drift over long times. We solve
the bias problem
by measuring only the modulation of the gravitational effect, $\Delta f_\mathrm{grav}$, instead of its total value. In practice, this means having two or several observational sessions at greatly
varying distances to the spacecraft.
Although the measured value of the modulation of $\Delta f_\mathrm{grav}$ is then free from the bulk of the bias, it still includes a contribution from the bias drift. The
latter, however, may be determined from a series  of observations at a constant geocentric distance. The drift measured this way (Fig.~\ref{fig:drift}), indeed, turns out to be non-negligible ($3.6\times10^{-14}$/day now), so that we must take it into account.

The second problem with Eq.~\eqref{eq:one-way} is associated with the nonrelativistic Doppler shift, $-\dot D/c$. 
Since the range rate error $\delta\dot D$ is $\sim$~2~mm/s \cite{zaslavsky-2016-la}, the error of the computed value of the Doppler
shift is $\delta(\dot D/c)\sim10^{-11}$, while $10^{-15}$ is required for achieving  $\delta\varepsilon\sim10^{-5}$.  
The first-order Doppler term, however, can be eliminated completely
(for a TS), or its magnitude reduced sufficiently (for a GRT), owing to the availability
of the 2-way ground--space--ground link (Fig.~\ref{fig:link-modes}). The 2-way link signal is sent by
a TS, received and phase-coherently retransmitted by the spacecraft, and finally
received again by a TS and/or a GRT. The frequency shift of the 2-way
link signal, for the simpler case of TS--space--TS propagation, is:
\begin{multline}
\Delta f_{2\mathrm{w}} =
f \bigg( 
 - 2{\dot D \over c} 
- {v_\mathrm{s}^2 - v_\mathrm{e}^2 \over c^2}
+ {|\mathbf v_\mathrm{s} - \mathbf v_\mathrm{e}|^2 \over c^2}
- 2 {\mathbf a_\mathrm{e} \cdot \mathbf n_{} \over c} \Delta t 
\\
+ 2{ (\mathbf v_\mathrm{s} \cdot \mathbf  n)^2
- (\mathbf v_\mathrm{e} \cdot \mathbf n) \cdot (\mathbf v_\mathrm{s} \cdot \mathbf n)
  \over c^2 }
\bigg)
\\
+ 2\Delta f_\mathrm{trop}
+ 2\Delta f_\mathrm{ion} 
+ O(v/c)^3,
\label{eq:two-way}
\end{multline}

\noindent where $\mathbf a_e$ is the ground station acceleration and $\Delta t$ is the
signal's light travel time \cite{vessot-levine-1979-grg}. (A physically similar but calculationally more complex equation
holds for the case of the 2-way link signal received by a nearby GRT.) Combining
the 1-way (\ref{eq:one-way}) and 2-way (\ref{eq:two-way}) frequency measurements, we obtain:

\begin{figure*}[t]
\includegraphics[width=\linewidth]{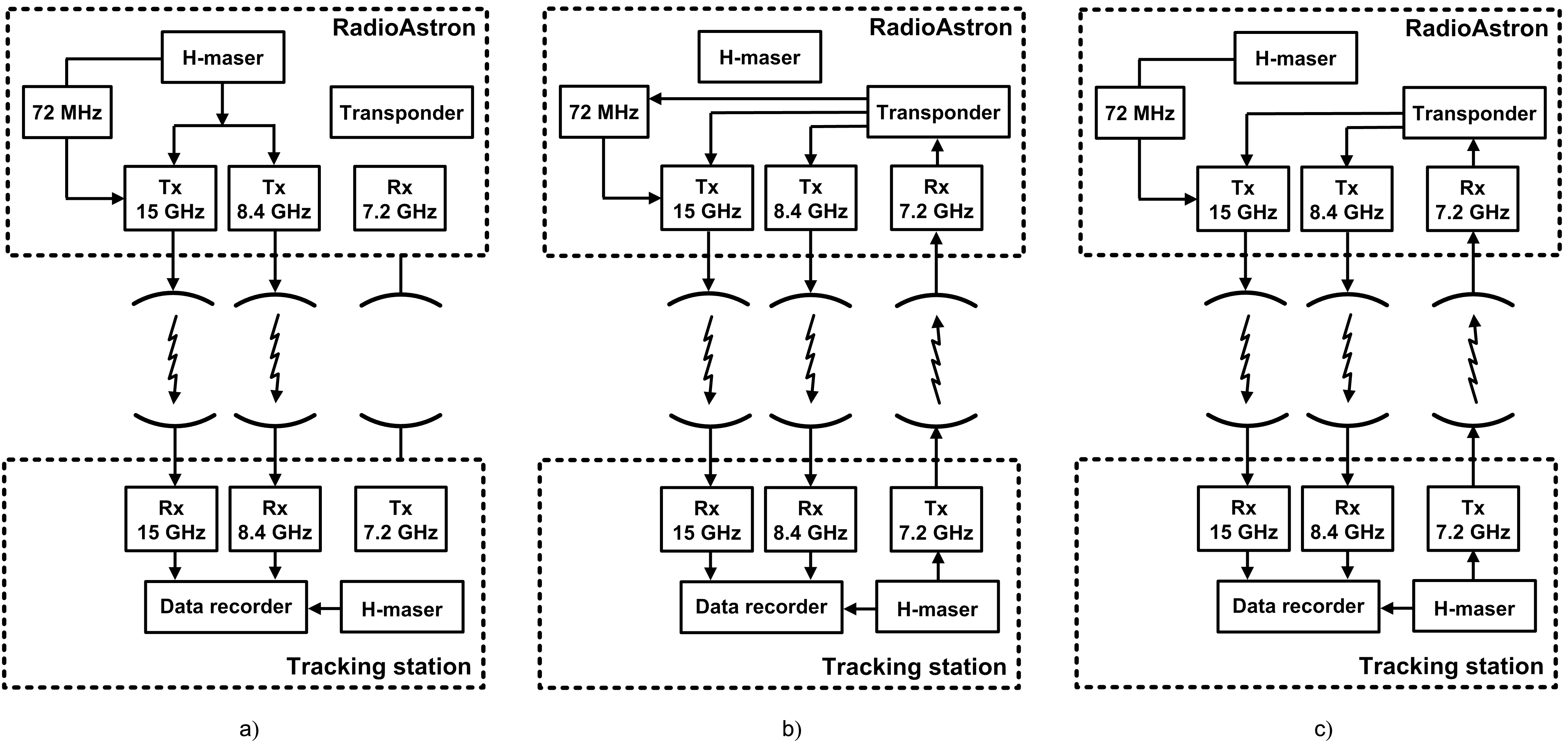}
\caption{Operation modes of the RadioAstron radio links: a) ``H-Maser'' (1-way); b) ``Coherent'' (2-way); c)~``Semi-Coherent'' (2-way).}
\label{fig:link-modes}
\end{figure*}

\begin{multline}
\Delta f_{1\mathrm{w}} - {\Delta f_{2\mathrm{w}} \over 2} 
=
\Delta f_\mathrm{grav}
+ f \left(
- {|\mathbf v_\mathrm{s} - \mathbf v_\mathrm{e}|^2 \over 2 c^2}
+ {\mathbf a_\mathrm{e} \cdot \mathbf n_{} \over c} \Delta t 
 \right) 
\\
+ \Delta f_0 
+ \Delta f_\mathrm{ion}^\mathrm{(res)} 
+ \Delta f_\mathrm{fine} 
+ O(v/c)^4,
\label{eq:gpa-compensation-scheme}
\end{multline}
where $\Delta f_\mathrm{ion}^\mathrm{(res)}$ is the residual ionospheric
shift (fully suppressed only for equal up- and downlink frequencies) and $\Delta f_\mathrm{fine}$ denotes several ``fine'' effects, such as those due to 
the relativistic kinematic terms of order $(v/c)^3$:
\begin{multline}
{\Delta f^{(3)} \over f}
=
{\mathbf n \cdot (\mathbf v_\mathrm{e} - \mathbf v_\mathrm{s}) \over c^3}
\left(
  \Delta U 
  - {|\mathbf v_\mathrm{s} - \mathbf v_\mathrm{e}|^2 \over 2}
  + ( \mathbf a_\mathrm{e} \cdot \mathbf n )  \, c \Delta t 
\right)
\\
+ 
{D\over c^3}
  \big( 
    -\mathbf v_\mathrm{s} \cdot \mathbf a_\mathrm{e}
    -( {\mathbf j}_\mathrm{e} \cdot \mathbf n  )  \, c \Delta t
    +2 \mathbf v_\mathrm{e} \cdot \mathbf a_\mathrm{e}
    +\mathbf v_\mathrm{e} \cdot \nabla U_\mathrm{e}
  \big),
\label{eq:third-order-kinematics}
\end{multline}
where ${\mathbf j}_\mathrm{e} = \dot {\mathbf a}_\mathrm{e}$ is the ground station jerk and $\nabla U_\mathrm{e}$ is the gradient of the gravitational potential at the ground station location (see Section 3 for other fine effects).
It is important to note that Eq.~\eqref{eq:gpa-compensation-scheme} is free from the nonrelativistic Doppler and tropospheric effects but retains the contribution of gravitation. The idea of the compensation scheme based on Eq.~\eqref{eq:gpa-compensation-scheme} was first realized in the GP-A mission, and the necessity of taking into account
third-order kinematic effects was noted in \cite{ashby-1998-ieee, blanchet-2001-aa}.
For RadioAstron, however, this scheme is not directly applicable because 1- and 2-way links cannot be operated simultaneously (Fig.~\ref{fig:link-modes}).
Nevertheless, two options for realizing the compensation scheme of Eq.~\eqref{eq:gpa-compensation-scheme} with RadioAstron have been devised.

The first option requires interleaving the 1-way  ``H-maser'' (Fig.~\ref{fig:link-modes}a) and 2-way ``Coherent''  (Fig.~\ref{fig:link-modes}b)  operation modes. The data
recorded by GRTs (and the TS) contain only one kind of signal at any given time.  
However, if the switching
cycle is short enough ($\sim$4~min at 8.4~GHz) we can interpolate the phases into the gaps with a corresponding frequency error of $\Delta f/f\sim4\times10^{-15}$. Thus we obtain simultaneous
frequency measurements of both kinds and can apply the compensation
scheme of Eq.~\eqref{eq:gpa-compensation-scheme} to them directly.
The approach based on interleaved measurements
does not rely on any features
of the signal spectrum and may be realized with telescopes equipped
either with 8.4 or 15 GHz receivers.

\begin{figure}[t]
\includegraphics[width=\linewidth]{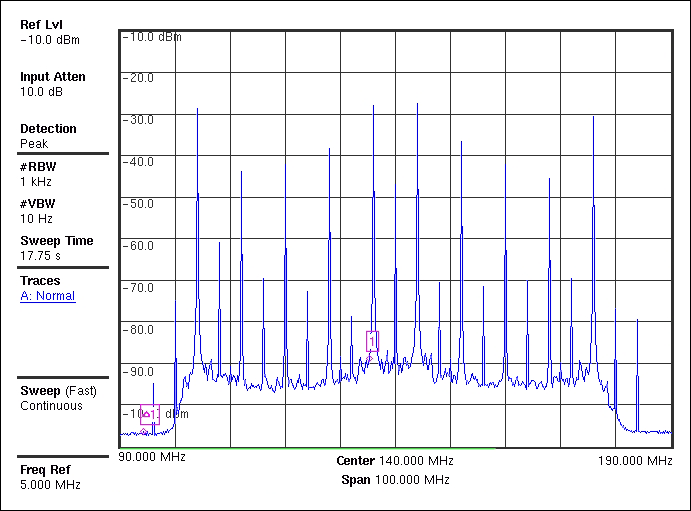}
\caption{Spectrum analyser screenshot, showing the signal spectrum of the 15 GHz data downlink in the ``Test-2'' mode of the on-board formatter.}
\label{fig:test-2}
\end{figure}

The second option for the Doppler compensation involves
recording the 15 GHz data downlink signal in the  ``Semi-Coherent'' mode of synchronization of the on-board scientific and radio equipment \cite{biriukov-2014-ar}, which
is a kind of half-way between the 1-way ``H-maser'' and 2-way ``Coherent'' modes. In this mode the  7.2 GHz
uplink tone, the 8.4 GHz downlink tone and the carrier of the 15 GHz data downlink
are phase-locked to the ground H-maser signal, while the modulation frequency
 (72 MHz) of the data downlink is phase-locked to the on-board H-maser signal
(Fig.~\ref{fig:link-modes}c).
This approach relies on the broadband ($\sim$1~GHz) nature of the
15 GHz signal modulated using quadrature phase-shift keying (QPSK)
and the
possibility of turning its spectrum into a comb-like form by  transmitting
a predefined periodic data sequence (Fig.~\ref{fig:test-2}).
As we have shown in \cite{biriukov-2014-ar},
different subtones of the resulting spectrum act like separate
links of the GP-A scheme and can be arranged in software postprocessing into a  combination
similar to that of Eq.~\eqref{eq:gpa-compensation-scheme}, which is free from the 1st-order Doppler
and tropospheric effects (the ionospheric term persists).

Despite some advantages of the second option from algorithmic and operational points of view, we give preference to the interleaved measurements approach as it provides for a larger number of participating GRTs
due to the larger ground footprint of the on-board antenna at 8.4 GHz and wider availability of 8.4 GHz receivers at GRTs.

\section{Data processing and fine effects}
\nopagebreak

The primary data for the experiment are the spacecraft signals at 8.4 and/or
15 GHz received and recorded at a ground station.  The majority of radio astronomy and geodetic radio telescopes are equipped
with H-maser standards and 8.4 GHz receivers, enabling them to take part in the experiment. Recording of the spacecraft signal is performed in
the ground H-maser timescale using standard VLBI back-end instrumentation. Initial data processing is based on the algorithms  developed originally for PRIDE (Planetary Radio Interferometry and Doppler Experiment) \cite{duev-2012-aa} for
recovering the phase of the received signal. Details of the
algorithm and software modifications required to process interleaved data will be given in an upcoming publication \cite{gusev-litvinov-rudenko-cqg-2017-to-be-published}. Here we briefly describe the approaches for correcting
the recovered signal phases for a number of  fine effects contributing to
the right-hand side of Eq.~\eqref{eq:gpa-compensation-scheme}:
\begin{itemize}

\item second- and third-order relativistic kinematic effects: computed
from the orbital data (velocity determination accuracy of 
$\delta v \sim$~2~mm/s \cite{zaslavsky-2016-la} is sufficient);

\item gravitational potential difference between the spacecraft and the ground
station: computed from the orbital data using the Earth gravitational potential
model \cite{pavlis-2008-jgr} (the position  error of  $\sim200$~m provided by radio ranging \cite{zaslavsky-2016-la}
is sufficient for distances $\gtrsim$~40,000~km, laser ranging required otherwise);

\item 
residual ionospheric frequency shift:
computed from 2-frequency measurements (8.4 and 15 GHz), ionospheric total electron content (TEC) maps \cite{hernandez-pajares-2009-jg} and mapping functions \cite{boehm-2006-jgr}, onsite GNSS receiver measurements;

\item 
frequency shift due to the tidal gravitational field of the Sun and Moon \cite{blanchet-2001-aa}: computed from the planetary and lunar ephemerides (JPL DE430);

\item 
phase center motion of the on-board and tracking station antennas: computed from
the orbital and housekeeping data \cite{moyer-2005-book};

\item
temperature dependence of the on-board H-maser: computed from the H-maser sensitivity
determined during ground tests and housekeeping data;

\item magnetic field dependence of the on-board H-maser frequency: computed from the H-maser sensitivity determined during ground tests, the magnetic
field model \cite{thebault-2015-eps} and the orbital data;

\item ground station motion due to solid Earth tides: computed from Earth models \cite{sovers-1998-rmp}.

\end{itemize}

After the gravitational frequency shift has been measured experimentally in a series of observations at various distances on a single orbit, we fit it against the gravitational
potential difference according to Eq.~\eqref{eq:main-freq}, thus obtaining a single measurement
of $\varepsilon$.  The accuracy of the result of 2--3 years
of planned data accumulation depends on the number of experiments performed
and their parameters. Based on the experiment error budget \cite{litvinov-mg14} and taking into account the observations performed
so far and those planned we expect the accuracy of the test to reach $\delta\varepsilon\sim10^{-5}$.

\begin{figure*}
\includegraphics[width=\textwidth]{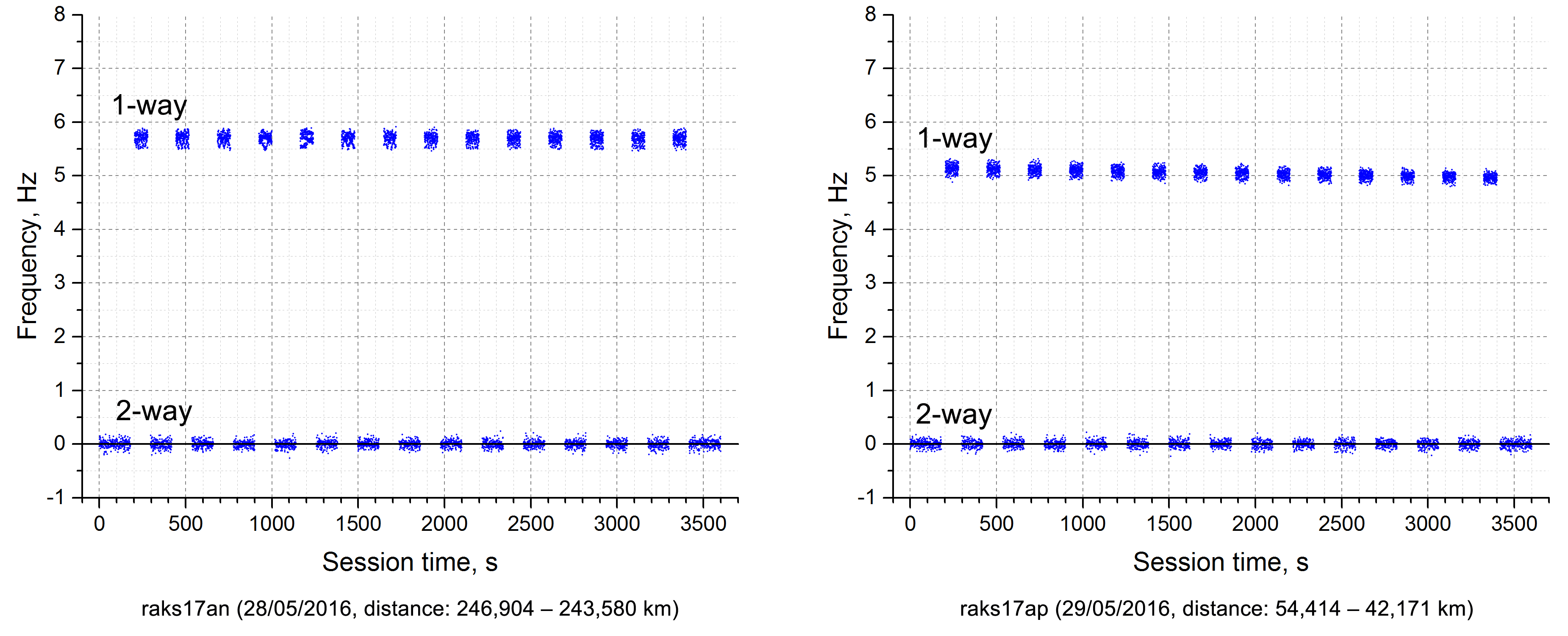}
\caption{Residual frequencies of the 1- and 2-way 8.4 GHz signals measured with the Onsala 20-m telescope. The 1-way frequency residuals are not corrected for the gravitational redshift. This makes the variation of the gravitational frequency shift between the two sessions clearly visible (varying from 5.69~Hz to 4.96~Hz).}
\label{fig:interleaved-residuals}
\end{figure*}

\section{Measurements}
\nopagebreak
 
The observations for the experiment are limited both by the technical constraints
of the RadioAstron satellite and competition for observational time with other science
projects of the RadioAstron mission. The technical constraints are: 
\begin{itemize}

\item 
the spacecraft's attitude limitations with respect to the Earth, Sun and Moon;

\item
the requirement that the spacecraft must be visible by the particular ground antenna;

\item
the requirement that the ground antenna must be within the on-board antenna's
ground footprint.

\end{itemize}

A total of 18 experiments have been performed
so far with the RadioAstron mission's Pushchino and Green Bank tracking
stations supported by several European VLBI Network telescopes (Effelsberg, Onsala, Svetloe, Wettzell, Yebes, Zelenchukskaya), the Robert C. Byrd Green Bank Telescope, and several Very Long Baseline Array antennas. 
Extensive tests have been performed with the radio telescopes and satellite laser ranging facilities of the Hartebeesthoek Radio Astronomy Observatory and AuScope VLBI Project's Yarragadee Observatory, which are to join the observations.
All experiments
were performed in the interleaved measurements mode (Fig.~\ref{fig:interleaved-residuals}) and consisted
of up to 4 observations, each $\sim 1$~hr long, distributed along
the orbit over  $\sim 20$--$50$~hr. Most observations were accompanied by satellite laser ranging to guarantee an orbit determination accuracy
at the cm-level \cite{pearlman-2002-asr}. 
The evaluation of our preliminary experimental results, which are consistent with $\varepsilon = 0$, will be published elsewhere.

\section{Conclusions}

The RadioAstron satellite, with its highly eccentric orbit and on-board H-maser frequency standard, is a unique space-borne laboratory for probing
the gravitational redshift effect, which constitutes a Local Position Invariance
test of the Einstein Equivalence
Principle. 
We have developed, and are implementing, a strategy for using the RadioAstron satellite to measure the gravitational redshift which takes into account the  limitations of the spacecraft. We should be able to measure the redshift to an accuracy of $\sim10^{-5}$, which is an order of magnitude better than the current best result of Gravity Probe A. Several measurements have already been obtained, and the data are being analyzed. Our preliminary results agree with the validity of the EEP. Some of the techniques we have developed, e.g. the particular realization of the Doppler compensation scheme, could be used for future space missions to test fundamental physics.

\section*{Acknowledgements}

The authors wish to thank the team of the Astro Space Center of the Lebedev Physical Institute, and especially its head Nikolay Kardashev, and the RadioAstron project scientist Yuri Y. Kovalev, for constant support in preparation and realization of the experiment. The authors also thank the referees for valuable comments.
Research for the RadioAstron gravitational redshift experiment is supported by the Russian Science Foundation grant 17-12-01488.
The RadioAstron project is led by the Astro Space Center of the Lebedev Physical Institute of the Russian Academy of Sciences and the Lavochkin Scientific and Production Association under a contract with the Russian Federal Space Agency, in collaboration with partner organizations in Russia and other countries.  
The European VLBI Network is a joint facility of independent European, African, Asian, and North American radio astronomy institutes. Scientific results from data presented in this publication are derived from the following EVN project codes: EL053, EL057.
The National Radio Astronomy Observatory is a facility of the National Science Foundation operated under cooperative agreement by Associated Universities, Inc.
The Long Baseline Observatory is a facility of the National Science Foundation operated under cooperative agreement by Associated Universities, Inc.
Hartebeesthoek Radio Astronomy Observatory is a facility of the South African National Research Foundation.
AuScope Ltd is funded under the National Collaborative Research
Infrastructure Strategy (NCRIS), an Australian Commonwealth Government
Programme.


\bibliography{pla}

\end{document}